\documentclass[a4paper,12pt]{article}

\usepackage[onehalfspacing]{setspace}
\usepackage{natbib}
\usepackage{url}
\usepackage{graphicx}

\title{Uncertainty in climate science and climate~policy}

\author{%
  Jonathan Rougier\thanks{Corresponding author.  Department of
    Mathematical Sciences, University of Bristol, University Walk,
    Bristol BS8 1TW, email
    \texttt{j.c.rougier@bristol.ac.uk}.}\\University of Bristol, UK\\[1ex]
  \and Michel Crucifix\\Universit\'{e} catholique de Louvain, Belgium}

\date{\small{}Draft copy.  File compiled \today, source \texttt{\jobname.tex}.}

\begin{document}

\maketitle

\section{Introduction}
\label{sec:introduction}

This essay, written by a statistician and a climate scientist,
describes our view of the gap that exists between current practice in
mainstream climate science, and the practical needs of policymakers
charged with exploring possible interventions in the context of
climate change.  By `mainstream' we mean the type of climate science
that dominates in universities and research centres, which we will
term `academic' climate science, in contrast to `policy' climate
science; aspects of this distinction will become clearer in what
follows.

In a nutshell, we do not think that academic climate science equips
climate scientists to be as helpful as they might be, when involved in
climate policy assessment.  Partly, we attribute this to an
over-investment in high resolution climate simulators, and partly to a
culture that is uncomfortable with the inherently subjective nature of
climate uncertainty.

In section~\ref{sec:modes} we discuss current practice in academic
climate science, in relation to the needs of policymakers.
Section~\ref{sec:uncertainty} addresses the aparently common
misconception (among climate scientists) that uncertainty is something
`out there' to be quantified, much like the strength of meridional
overturning circulation.  Section~\ref{sec:riskManager}, the heart of
the essay, addresses the core needs of the policymaker, and focuses
on three strictures for the climate scientist wanting to help her:
answer the question, own the judgement, and be coherent.
Section~\ref{sec:reflection} concludes with a brief reflection.

We have taken the opportunity in this essay to be a little more
polemical than we might be in an academic paper, and maybe a little
more exuberent in our expressions.  We have also ignored the technical
details of practical climate science, something we are both involved
in day-to-day, choosing instead to look at the larger picture.  We
believe that our observations are valid more widely than just climate
science; for example many of them would apply with little modification
in many areas of natural hazards, and in radiological or
eco\-toxicological risk assessment \citep{rougier13sappur}.  But they
seem most pertinent in climate science, which outstrips the other
areas in terms of funding.  For example, the UK's Natural Environment
Research Council (NERC), whose vision is to ``advance knowledge and
understanding of the Earth and its environments to help secure a
sustainable future for the planet and its people'', allocates $40\%$
of its science budget to climate science and earth system science
(NERC Annual Report and Accounts 2010--11, p.~40).

\section{Different modes of climate science}
\label{sec:modes}

For our purposes, the telling feature of climate science is that it
gained much of its momentum in the era before climate change became a
pressing societal concern.  Consequently, when policymakers turned to
climate science for advice, they encountered a well-developed academic
field whose focus was more towards explanation than prediction.
Explanation, in this context, is verifying that observable
regularities in the climate system are emergent properties of the
basic physics.  Largely this is through the interplay between
observation and dynamical climate simulation.  As the resolution of
climate simulators increases, more observed regularities fall into the
`explained' category.  The El Ni\~{n}o Southern Oscillation (ENSO) is
getting closer to falling into this category, for example
\citep{guilyardi09}.

Thus for investment, the dominant vector in academic climate science
has been to improve the spatial and temporal resolution of the solvers
in climate simulators.  Supporting evidence can be found in
meteorology.  It is argued that one of the contributory factors to
measurable improvements in weather forecasting over the last thirty
years is higher-resolution solvers, although the quantification of
this is confounded by simultaneous improvements in understanding the
physics, in the amount of data available for calibration, and in
techniques for data assimilation \citep[ch.~1]{kalnay02}.  Setting
these confounders aside, it seems natural to assert that higher
resolution solvers will lead to better climate simulators.  And
indeed, we would not deny this, but we would also question whether in
fact it is resolution that is limiting the fidelity of climate
simulators.

The reason that we are suspicious of arguments about climate founded
on experiences in meteorology is the presence of biological and
chemical processes in the earth system that operate on climate policy
but not weather time-scales.  We believe that the acknowledgement of
biogeochemistry as a full part of the climate system distinguishes the
true climate scientist from the converted meteorologist.  Our lack of
understanding of climate's critical ecosystems mocks the precision
with which we can write down and approximate the Navier-Stokes
equations.  The problem is, though, that putting ecosystems into a
climate simulator is a huge challenge, and progress is difficult to
quantify.  It introduces \emph{more} uncertain parameters, and, by
replacing prescribed fields with time-evolving fields, it can actually
make the performance of the simulator worse, until tuning is
successfully completed (and there is no guarantee of success).
\citet{newman11} provides a short and readable account of the
difficulties of biology, in comparison to physics.

On the other hand, spending money on higher resolution solvers
requires \emph{fewer} parameterisations of sub-grid-scale processes,
and so reduces the challenge of tuning.  This activity has a
well-documented provenance, and a clear motivation within a coherent
science plan. And we cannot resist pointing out another immediate
benefit: one can show the funder a more realistic looking ocean
simulation (``Now at $0.5^\circ$ resolution!'')---although in fact
resolutions as high as $0.1^\circ$ do not fool experienced
oceanographers.  But while this push to higher resolutions is natural
for meteorology, with its forecast horizon measured in days, for
climate we fear that it blurs the distinction between what one
\emph{can} simulate, and what one \emph{ought} to simulate for policy
purposes.

So how might the investment be directed differently?  For climate
policy it is necessary to enumerate what might happen under different
climate interventions: do nothing, monetise carbon, regulation for
contraction and convergence, geo-engineering, and so on.  And each of
these interventions must be evaluated for a range of scenarios that
capture future uncertainty about technology, economics, and
demographics.  For each pair of intervention and scenario there is a
range of possible outcomes, which represent our uncertainty about
future climate.  Uncertainty here is `total uncertainty': only the
intervention and the scenario are specified---the policymaker does not
have the luxury of being able to pick and choose which uncertainties
are incorporated and which are ignored.

\begin{figure}
  \centering
  \includegraphics{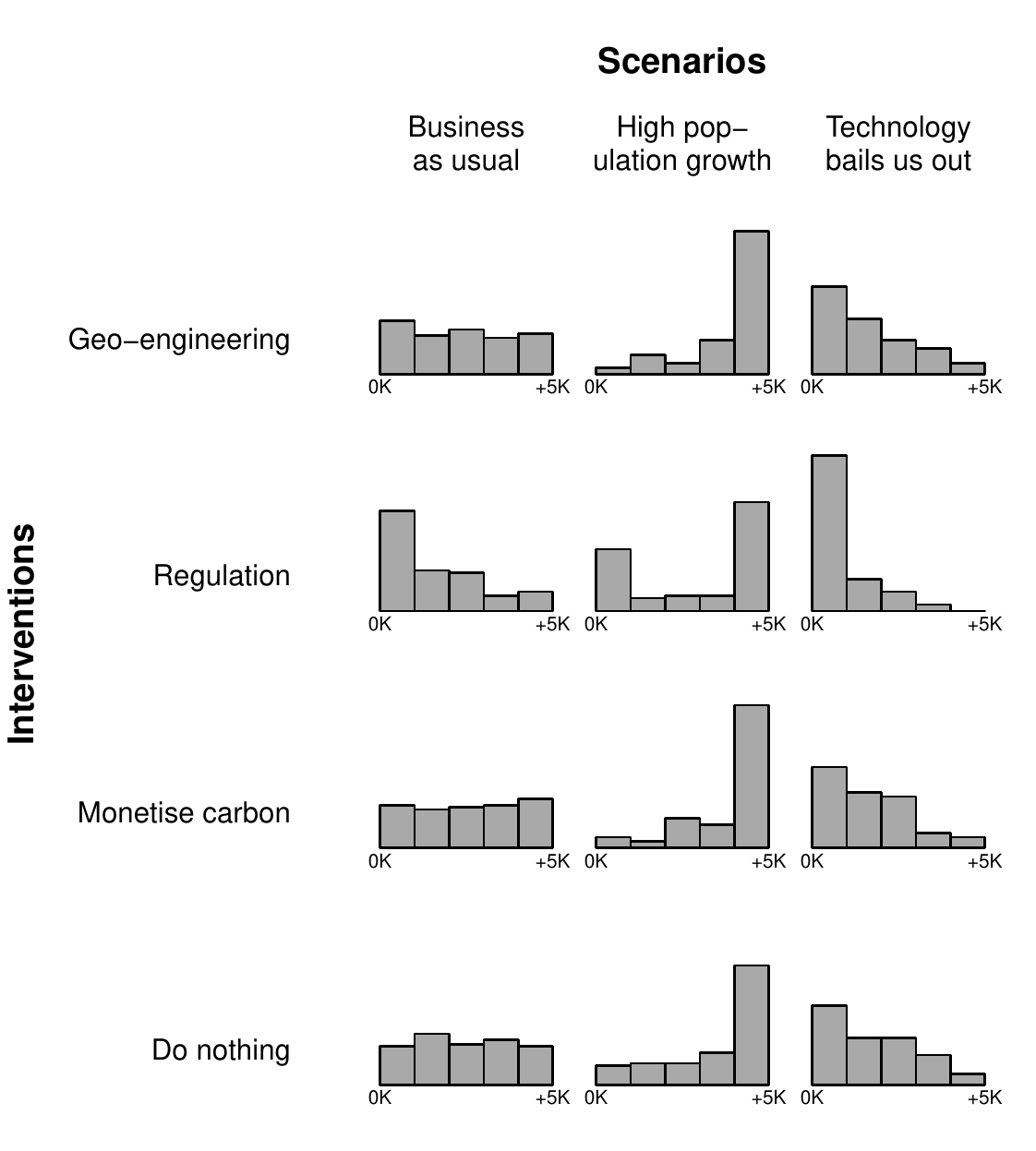}
  \caption{Policy tableau, showing the effect of different possible
    interventions under different scenarios.  These frequency
    histograms might in this case measure simulated global warming by
    2100 under different not-implausible simulator configurations, but
    more generally they would measures losses, inferred from simulated
    distributions for weather in 2100. Please note that these
    histograms are \emph{completely fictitious!}}
  \label{fig:tableau}
\end{figure}

Internal variability, part of the natural variability of the climate
system, can be estimated from high-resolution simulators, but it is
only a tiny part of total uncertainty.  Over centurial scales, it is
negligible compared to our combined uncertainty of the behaviour of
the ice-sheets, and the marine and terrestrial biosphere.  This
uncertainty can be assessed with the assistance of climate simulators,
if it is possible to run them repeatedly under different
configurations of the simulator parameters and modules, where these
configurations attempt to span the range of not-implausible climate
system behaviours.  To construct a tableau such as the one in
Figure~\ref{fig:tableau} will require a minimum of $4 \times 3 \times
100 \times 90$ models-years of simulation, say 120,000 model-years,
including spin-up.  The $100$ is the number of different simulator
configurations that might be tried, and the $90$ is the number of
years until 2100.  Of course, $100$ is woefully small for the number
of configurations.  There are more than one hundred uncertain
parameters in a high-resolution climate simulator \citep{murphy04}.
Admittedly only some of these will turn out to be important but we
cannot rule out interactions among the parameters.  There is a
well-developed statistical field for this type of analysis, see, e.g.,
\citet{santner03}.

Note that this is a designed experiment, deliberately constructed to
be informative about uncertainty.  It is completely different from
assembling an \textit{ad hoc} collection of simulator runs, such as
the CMIP3 or CMIP5 multimodel ensembles, in the same way that a
carefully stratified sample of $100$ people is far more informative
about a population than simply selecting the next $100$ people that
pass a particular lamp-post.  In the absence of designed experiments,
though, climate scientists who want to assess uncertainty will have to
use the \textit{ad hoc} ensemble.  The various types and uses of
currently-available ensembles of climate simulator runs are reviewed
in \citet{parker10} and \cite{murphy11}.

So what is the status of these policy-relevant designed experiments?
Current `IPCC class' simulators (with a solver resolution of about
$1^\circ$) run at about 100 model-years per month of wall-clock time.
So starting now, an experiment to assess uncertainty in 2100 for
policy purposes will be finished in about 100 years, if it is
performed at one research centre.  But this might be reduced to 10
years if the runs were shared out across all centres, or even less
factoring in faster computers and no increase in resolution.  Thus
these IPCC class simulators could be very helpful for assessing
uncertainty and supporting policymakers, but this requires a cap on
solver resolution, and careful coordination across research centres.
In contrast, the current uncoordinated approach, with its apparent
commitment to spending CPU cycles on a few runs of high-resolution
climate simulators, will force climate scientists in 2020 to base
their future climate assessments on \textit{ad hoc} ensembles.

\section{The nature of uncertainty about climate}
\label{sec:uncertainty}

In this paper we confine our discussion of climate uncertainty
quantification to the assessment of probabilities.  There are, of
course, several interpretations of probability.  L.J. Savage wrote of
``dozens'' of different interpretations of probability
\citet[p.~2]{savage72}, and he focused on three main strands: the
Objective (or Frequentist), the Personalistic, and the Necessary.
This tripartite classification is widely accepted among statisticians,
and discussed, with embellishments, in the initial chapters of
\citet{walley91} and \citet{lad96}.  Not to be outdone,
\citet{hajek12} notes that philosophers of probability now have six
leading interpretations of probability.

Of all of these interpretations, however, we contend that only the
Personalistic interpretation can capture the `total uncertainty'
inherent in the assessment of climate policy.  Our uncertainty about
future climate is predominantly \emph{epistemic} uncertainty---the
uncertainty that follows from limitations in knowledge and resources.
The hallmark of epistemic uncertainty is that it could, in principle,
be reduced with further introspection, or further experiments.  As one
of the key drivers of research investment in climate science is to
reduce uncertainty, this epistemic interpretation of `total
uncertainty' must be uncontentious.  It rules out the Objective
(classical, frequency, propensity) interpretation, and leaves us with
Personalistic and Necessary (also termed logical) interpretations.

The Necessary interpretation asserts that there are principles of
reasoning that extend Boolean logic to uncertainty, and that these
principles are in fact the calculus of probability and Bayesian
conditioning.  This interpretation is formally attractive, but invokes
additional principles to `fill in' those initial probabilities that
are mandated by conditioning---which are generally referred to as
`prior' probabilities in a Bayesian context.  These are to be based on
self-evident properties of the inference, such as symmetries.
Examples are discussed in \citet{jaynes03}; see, for example, his
elegant resolution of Bertrand's problem (sec.~12.4.4).  However, it
is hard to know how one might discover and apply these properties in
an assessment of, say, the maximum height of the water in the Thames
Estuary in 2100.  Thus, starting with Frank~Ramsey, and finding
eloquent champions in Bruno de~Finetti and L.J.~Savage, among others,
the Personalistic interpretation has provided an operational
subjective definition of probability, in terms of betting rates
\citep[see, e.g.][]{ramsey31,finetti37english,savage72,savage62}.
De~Finetti's late writings are both subtle and discursive;
\citet{lad96} attempts to corral them.

Not everyone will find the Personalistic definition of probability
compelling.  But at least it provides a very clear answer to the
question ``What do You mean when You state that $\Pr(A) = p$?''  A
brief answer is that, if betting for a small amount of money, such as
$\pounds 1$, You would be agreeable to staking up to $\pounds p$ in a
gamble to receive $\pounds 1$ if $A$ turns out to be true and nothing
if $A$ turns out to be false. There are other operationalisations as
well, which are very similar but not psychologically equivalent; see,
e.g., the discussion in \citet[sec.~2.2]{gw07}.  Our view is that an
operationalisation of Personalistic probability is highly desirable,
and a useful thing to fall back on, but not in itself the yardstick by
which all probabilities are assessed.  But, if someone provides a
probability $p$ for a proposition $A$, it might be a good idea to ask
him if he would be prepared to bet $\pounds p$ on $A$ being true: the
answer could be very revealing.

However, many physical scientists seem to be very uncomfortable with
the twin notions that uncertainty is subjective (i.e.\ it is a
property of the mind), and that probabilities are expressions of
personal inclinations to act in certain ways.  At least part of the
problem concerns the use of the word `subjective', about which the
first author has written before \citep[sec.~2]{rougier07cc}.  This
word is clearly inflammatory.  We suggest that some scientists have
confused the Mertonian scientific norm of `disinterestedness' with the
notion of `objectivity', and then taken subjectivity to be the
antithesis of objectivity, and thus to be avoided at all costs.
L.J.~Savage was sensitive to this confusion and hence favoured
`Personalistic'.  De~Finetti strongly favoured `subjective', about
which \citet[p.~76, footnote~1]{jeffrey04} commented on ``the lifelong
pleasure that de~Finetti found in being seen to give the finger to the
establishment''.

Confusion about `subjectivity' is just a digression, though.  What is
abundantly clear is that climate scientists are not ready to accept
that climate uncertainties are Personalistic.  Their every reference
to `\emph{the} uncertainty' commits an error which the physicist
E.T.~Jaynes called the `mind projection fallacy':
\begin{quote}
  an almost universal tendency to disguise epistemological statements
  by putting them into a grammatical form which suggests to the unwary
  an ontological statement.  To interpret the first kind of statement
  in the ontological sense is to assert that one's own private
  thoughts and sensations are realities existing externally in Nature.
  \citep[p.~22]{jaynes03}.
\end{quote}
Jaynes is an example of a physicist who embraced the essential
subjectivity of uncertainty: he advocated the Necessary
interpretation, plus the additional principle of maximising Shannon
entropy to extend limited judgements to probabilities.
\citet{paris94} provides a detailed assessment of the properties of
this entropy-maximising approach, among others.

One very stealthy manifestation of the Mind Projection Fallacy is the
substitution of `assumptions' for `judgements' when discussing
uncertainty.  Assumptions typically refer to simplifications we assert
about the system itself.  It is perfectly acceptable to \emph{assume}
that, for example, the hydrostatic approximation holds: this is a
statement that actual ocean behaves a lot like a slightly different
ocean that is much simpler to analyse.  You cannot \emph{assume},
though, that the maximum water level in the Thames Estuary in 2100 has
a Gaussian distribution.  Instead, You may judge it appropriate to
represent Your uncertainty about the maximum water level with a
Gaussian distribution.  This is rather wordy, unfortunately, which is
perhaps why it is so easy to lapse in this way.

Consider the uncertainty assessment guidelines for the forthcoming
IPCC report \citep{mastrandrea10}.  Nowhere in the guidelines was it
thought necessary to define `probability'.  Either the authors of the
guidelines were not aware that this concept was amendable to several
different interpretations, or that they were aware of this, and
decided against bringing it out into the open.  One can imagine, for
example, that an opening statement of the form ``In the context of
 climate prediction, probability is an expression of subjective
uncertainty and it can be quantified with reference to a subject's
betting behaviour'' would have caused great consternation---so much
the better!

We can hardly suppose that the omission of a definition for the key
concept in such an important and high-profile document was made in
ignorance.  And yet the mind projection fallacy is in evidence
throughout.  It looks as though the authors have deliberately chosen
\emph{not} to acknowledge the essential subjectivity of climate
uncertainty, and to suppress linguistic usage that would indicate
otherwise.  This should be termed `monster denial' in the taxonomy of
\citet{curry11}.  Choosing not to rock the boat is convenient for
academic climate scientists.  But it makes life difficult for
policymakers, who are tasked with turning uncertainties into actions.
For policymakers, the meaning of `$\Pr(A) = p$' is of paramount
importance, and they need to know if ten different climate scientists
mean it ten different ways.
% , especially if some of those ways are
% hardly defensible.  (Our personal favourite: the `many worlds'
% interpretation which equates probabilities with limiting frequencies
% in a collective of hypothetical `similar but not identical' worlds.
% It is fascinating when physicists go metaphysical.)

\section{The risk manager's point of view}
\label{sec:riskManager}

In any discussion of uncertainty and policy it is helpful to label the
key players \citep[ch.~1]{smith10}.  Conventionally, the person who
selects the intervention is the \emph{risk manager}, who represents a
particular set of stakeholders.  These stakeholders, who are funding
the risk manager, and will also fund the intervention that she
selects, will appoint an \emph{auditor}, whom the risk manager must
satisfy.  This framework, of a risk manager who must satisfy an
auditor, is a simple way to abstract from the complexities of any
particular decision.  It emphasises that the risk manager is an agent
who must defend her selection, and this has important consequences for
the way in which she acts.

The risk manager is surely uncertain about future climate, and its
implications.  For concreteness, suppose that her concern is about the
maximum height of water in the Thames Estuary in the year 2100.  If
asked, she might say, ``Really, I've no idea, perhaps not lower than
today's value, and not more than two metres higher.''  But she is not
obliged to make such an assessment in isolation: she can consult an
\emph{expert}.  Put simply, her expert is someone whose judgements she
accepts as her own \citep[see][sec.~6.3 for a discussion]{lad96}.  So
one task of the risk manager is to select her expert, and she must do
this in such a way that the auditor is satisfied with the selection
process, and with the elicitation process.  When seen from the other
side, it follows that scientists who want to be involved in climate
policy are competing with each other to be selected as one of the risk
manager's experts.  Therefore they must demonstrate their grasp of the
risk manager's needs.  Likewise, for climate scientists who are
competing for policy-tagged funding.

We highlight the following three risk managers' needs, as posing
particular challenges for academic climate scientists.

\subsection{Answer the question}
\label{sec:question}

As already discussed, the risk manager needs an assessment of `total
uncertainty'.  It can be difficult for the climate scientist to assess
his total uncertainty about future climate because of academic climate
science's focus on consuming CPU cycles in higher-resolution solvers,
rather than designed replications across alternative not-implausible
configurations of simulator parameters and modules.  This leaves the
willing-to-engage climate scientist ill-equipped to answer questions
about ranges for future climate values, because he has nothing other
than intuition to guide him on the consequence of the limitations in
our knowledge.  Unfortunately, his intuition may be tentative at best
when reasoning about a dynamical system as complex as the climate
system, on centurial timescales.

In this case, the climate scientist may end up specifying very wide
intervals which, although honest, do not advance the risk manager
because they swamp any `treatment effect' that might arise from
different choices of intervention.  This honest climate scientist may
well be passed over in favour of other experts who advertise their
smaller uncertainty as a putative measure of their superior expertise.
This type of competition is extensively discussed in
\citet{tetlock05}, in the context of political and economic
forecasting, and the parallels with climate forecasting seem very
strong.

How to make the uncertainties smaller?  One way is to qualify them
with conditions.  If these conditions are specified in the question,
then of course this is fine.  If the risk manager, for example, wants
to know about the height of the water in the Thames Estuary under the
`Technology bails us out' scenario, then in it goes.  But everything
else is suspect.  Sometimes the qualification is overt, for example
one hears ``assuming that the simulator is correct'' quite frequently
in verbal presentations, or perceives the presenter sliding into this
mindset.  This is so obviously a fallacy that he might as well have
said ``assuming that the currency of the US is the jam doughnut''.
The risk manager would be justified in treating such an assessment as
meaningless.  After all, if the climate scientist is not himself
prepared to assess the limitations of the simulator, then what hope is
there for the risk manager?

As \citet{tetlock05} documents, though, often the qualifications are
implicit, and only ever appear at the point where the judgement has
been shown to be wrong, e.g.\ ``Well, of course I was assuming that
the simulator was correct''.  The risk manager is not going to be able
to winkle out all of these implicit conditions at the start of the
process, but other climate scientists might be able to.  Thus the
elicitation process must be very carefully structured to ensure that,
by the time that the experts finally deliver their probabilities, as
many as possible of the implicit qualifications have been exposed and
undone.  This usually involves a carefully facilitated group
elicitation, typically extending over several days.  Interestingly,
\citeauthor{tetlock05} did not use group elicitations in his study,
but they are standard in environmental science areas such as natural
hazards; see, e.g., \citet{cooke00}, \citet{aspinall10}, or
\citet{aspinall13}.

Scientists working in climate, and philosophers too we expect, often
receive requests to complete on-line surveys about future climate.
These surveys are desperately flawed by responses missing `not at
random'. But even were they not, their results ought to be treated
with great circumspection, given the experience in natural hazards of
how much difference a careful group elicitation can make, in comparing
experts' probabilities at the start and at the finish of the process.

\subsection{Own the judgement}
\label{sec:ownership}

This is in fact another type of qualification, where the climate
scientist does not present his own judgement, but someone else's.  A
classic example would be ``according to the recent IPCC report''.  As
far as the climate scientist is concerned, these qualified uncertainty
assessments are consequence-free, and they ought to be judged by the
risk manager as worthless, since nothing is staked.  % Many people would
% appear to disagree with this; \citet{oppenheimer??}, for example.  But
% it is important to be careful about the context.
% \citetauthor{oppenheimer??} was referring to climate science
% communication, particularly in areas outside one's primary expertise.

The IPCC reports are valuable sources of information, but no~one owns
the judgements in them.  Only a very na\"{\i}ve risk manager would
take the IPCC assessment reports as their expert, rather than
consulting a climate scientist, who had read the reports, and also
knew about the culture of climate science, and about the IPCC process.
This is not to denigrate the IPCC, but simply to be appropriately
realistic about its sociological and political complexities, in the
face of the very practical needs of the risk manager.  These
complexities are well-recognised, and a decision by the risk manager
to adopt the IPCC reports as her expert can hardly be blame-free.  As
a marketing ploy, the decision to buy IBM computers was said to be
blame-free in the 1970s and 80s: ``nobody ever got fired for buying
IBM equipment''---how hollow that sounds now!

The challenge with owning the judgement in climate science is the
complexity of the science itself.  There are three main avenues for
developing quantitative insights about future climate: (i)~computer
simulation, (ii)~contemporary data collected mainly from field
stations, ocean sondes, and satellites, but also slightly older data
from ships' log-books, and (iii)~pal\ae{}o\-climate reconstruction
from archives such as ice and sediment cores, speliothems, boreholes,
and tree-rings.  Each of these is a massive exercise in its own right,
involving large teams of people, large amounts of equipment, and
substantial numerical processing.  Judgements about future climate at
high spatial and temporal resolution come mainly from computer
simulation, but one must not forget that these simulators have been
tuned and critiqued against contemporary data and, increasingly,
pal\ae{}o\-climate reconstructions.

Wherever there is a high degree of scientific complexity, there is a
large opportunity for human error.  With computer simulation, an
often-overlooked opportunity for error is the wrapping of the
computational core for a specific task; for example, performing a
time-slice experiment for the Mid-Holocene at a particular combination
of simulator parameter values.  Whereas the computational core of the
simulator is used time and again, and one might hope that large errors
will have been picked up and corrected and committed back to the
repository, the wrapper is often used only once.  It tends to be
poorly documented, often existing as a loose collection of scripts
which are passed around from one scientist to another.  It is easy to
load the wrong initialisation file or boundary file, and also easy to
extract the wrong summary values from the gigabytes of simulator
output.  `Easy' in this case equates to `if you have done an
experiment like this, you will be aware of at least one mistake that
you made, spotted, and corrected'.  The correction of this type of
mistake can take weeks of effort, as it is tracked backwards from the
alarming simulator output to its source in the underlying code.

% So owning a judgment about future climate based solely on a
% large-scale computer simulation involves a high degree of trust, and
% the same must be true for the other sources of quantitative
% information.  Naturally, one respects one's colleagues, but does one
% entrust one's reputation to them?
At the other end of the modelling spectrum, there are phenomenological
models of low-dimensional properties of climate and its impacts.  See,
for example, \citet{crucifix12}, who surveys dynamical models of
glacial cycles, or \citet{lorenz12}, who study the welfare value of
reducing uncertainty, notably in the presence of a climate tipping
point.  There are several advantages to such models.  First, they are
small enough to be coded by the scientist himself, and can be
carefully checked for code errors.  Thus the scientist can himself be
fairly sure that the interesting result from his simulator is not an
artifact of a mistake in the programming.  Second, they are often
tractable enough to permit a formal analysis of their properties.  For
example, they might be qualitatively classified by type, or explicitly
optimised, or might include intentional agents who perform sequences
of optimisations (such as risk managers).  Third, they are quick
enough to execute that they can be run for millions of model years.
Hence the scientist can use replications to assimilate measurements
(including tuning the parameters) and to assess uncertainty, both
within a statistical framework \citep[e.g., using the sequential
approach of][]{andrieu10}.

Of course, `big modellers' will be scornful of the limited physics
(biology, chemistry, economics, \textit{etc.})\ that these
phenomenological models contain, although they must be somewhat
chastened by the inability of their simulators to conclusively
outperform simple statistical procedures in tasks such as ENSO
prediction \citep{barnston12}.  But the real issue is one of
ownership.  A single climate scientist cannot own an artifact as
complex as a large-scale climate simulator, and it is very hard for
him to make a quantitative assessment of the uncertainty that is
engendered by its limitations.  We advocate spending resources on
designed experiments to support the climate scientist in this
assessment, but we also note that a scientist \emph{can} own a
phenomenological model, and the judgements that follow from its use.

% But big modellers need to convince
% other climate scientists and the risk manager that they have
% scrutinised the million or so lines of code with the same thoroughness
% that  can easily be applied to a few hundred.  One option, which does not
% guarantee scrutiny but at least makes it feasible, is to make climate
% simulators open source, which is the case for the Community Earth
% System Model (\url{http://www.cesm.ucar.edu/}).  This simulator is
% described in \citet{gent11}, which is an excellent summary of the
% issues involved in commissioning a large-scale climate simulator, and
% the choices made by climate scientists during this process.

\subsection{Be coherent}
\label{sec:coherent}

\citet[p.~7]{tetlock05} has a similar requirement.  In this context,
`coherent' has a technical meaning, which is to say, `don't make
egregious mistakes in probabilistic reasoning'.  This needs to be
said, because it is more honour'd in the breach than the observance.

For example, \citet{gigerenzer03} provides a vivid account of how
doctors, who ought to be good at uncertainty assessment, often
struggle with even elementary probability calculations, and how this
compromises the notion of informed consent to medical procedures.  As
another example, the `$P$-value fallacy'---inferring that the null
hypothesis is false because the $P$-value is small---is endemic in
applied statistics \citep[see, e.g.,][]{goodman99a,ioannidis05}.  It
is very similar to the Prosecutor's fallacy in Law \citep[see,
e.g.,][ch.~9]{gigerenzer03}. These fallacies serve to remind us that
people are not very good when reasoning about uncertainty, and that
they can easily be mislead by fallacious arguments (that violate the
probability calculus), sometimes intentionally.

%  The $P$-value fallacy is not just a
% theoretical concern: it is a contributory factor in the alarmingly low
% rates of reproducibility in medical science \citep{ioannidis05}.
% There is no reason to think that medical science is less sophisticated
% in its inferences than climate science; in fact quite the opposite.

\citet[ch.~4]{tetlock05} also notes another aspect of coherence, which
is to appropriately update opinions in the light of new information.
He emphasises the use of Bayes's Theorem, and demonstrates that his
experts did not make the full adjustment that was indicated by
Bayesian conditioning.  While there are psychological explanations for
under-adjustment, we would also note that the probability calculus and
Bayesian conditioning is only a \emph{model} for reasoning about
uncertainty, and not the \textit{sine qua non}.

Probabilistic inference owes its power to the unreasonable demands of
its axioms, notably the need to quantify an additive (probability)
measure on a sufficiently rich field of propositions.  This point was
very clearly expressed by \citet[notably sec.~2.5]{savage72}, in his
contrast between the small world in which one assesses probabilities
and performs calculations, and the grand world in which one makes
choices.  He writes ``I am unable to formulate criteria for selecting
these small worlds and indeed believe that their selection may be a
matter of judgment and experience about which it is impossible to
enunciate complete and sharply defined general principles \dots\ On
the other hand it is an operation in which we all necessarily have
much experience, and one in which there is in practice considerable
agreement'' (pp.~16-17).

A similar point is made by \citet[ch.~3]{howson06}, who defend precise
probabilities as a \emph{model} for reasoning against more complex
variants in terms of ``the explanatory and informational dividends
obtained from their use within simplifying models of
\textit{uncertain} inference'' (p.~62, original emphasis).
\citeauthor{howson06} present an instructive analogy with deductive
logic, whose poor representation of implication requires that we use
it thoughtfully when reasoning about propositions that are either true
or false (p.~72).  Thus in reasoning about uncertainty, grand world
probabilities will be informed by small world calculations such as
Bayesian conditioning, but need not be synonymous with them.  The
Temporal Sure Preference condition of \citet{goldstein97} provides one
way to connect these two worlds \citep[see also][sec.~3.5]{gw07}.

So, for climate scientists, and the risk managers they are hoping to
impress, the moral of \emph{be coherent} is that (i)~it is very easy
to make mistakes when reasoning about uncertainty, that (ii)~strict
adherence to the rules of the probability calculus (and perhaps the
assistance of a professional statistician) will minimise these, and
that (iii)~although probability calculations are highly informative,
no one should be overly impressed by an uncertainty assessment that is
a precise implementation of fully probabilistic Bayesian
conditioning---one would expect this to be simplistic.

\section{Reflection}
\label{sec:reflection}

Suppose that you were one of a group of climate scientists, interested
in playing an active role in climate policy, and able to meet the
three strictures outlined in section~\ref{sec:riskManager}.  You have
all embraced subjective uncertainty, and have been summoned,
willingly, to a carefully facilitated expert elicitation session.
After two intense but interesting days your $95\%$ equi-tailed
credible interval for the maximum height of water in the Thames
Estuary in 2100 is $0.5$m to $2.75$m higher than today.  This is wider
than your initial interval, as you came to realise, during the
elicitation process, that there were uncertainties which you had not
taken into account.

Suppose that this has recently happened, and you are reflecting on the
process, and wondering what information might have made a large
difference to your uncertainty assessment, and that of your fellow
experts.  In particular, you imagine being summoned back in the year
2020, to re-assess your uncertainties in the light of eight years of
climate science progress.  Would you be saying to yourself, ``Yes,
what I really need is an \emph{ad hoc} ensemble of about $30$
high-resolution simulator runs, slightly higher than today's
resolution.''  Let's hope so, because right now, that's what you are
going to get.

But we think you'd be saying, ``What I need is a designed ensemble,
constructed to explore the range of possible climate outcomes, through
systematically varying those features of the climate simulator that
are currently ill-constrained, such as the simulator parameters, and
by trying out alternative modules with qualitatively different
characteristics.''  Obviously, you'd prefer higher resolution to the
current resolution, but you don't see squeezing another $0.25^\circ$
out of the solver as worth sacrificing all the potential for exploring
uncertainty inherent in our limited knowledge of the earth system's
dynamics, and its critical ecosystems.  We'd like to see at least one
of the large climate modelling centres commit to providing this
information by 2020, on their current simulator, operating at a
resolution that permits hundreds of simulator runs per scenario (a
resolution of about $2^\circ$, we hazard).  Research funders have the
power to make this happen, but for some reason they have not yet
perceived the need.

\singlespacing\small
%\bibliography{statistics,climate,dynamical,ComputerExperiment,palaeo}

\end{document}